# Probing dynamic behavior of electric fields and band diagrams in complex semiconductor heterostructures

Yury Turkulets and Ilan Shalish*

*Ben Gurion University of the Negev, Beer Sheva, Israel*

Modern bandgap engineered electronic devices are typically made of multi-semiconductor multi-layer heterostructures that pose a major challenge to silicon-era characterization methods. As a result, contemporary bandgap engineering relies mostly on simulated band structures that are hardly ever verified experimentally. Here, we present a method that experimentally evaluates bandgap, band offsets, and electric fields, in complex multi-semiconductor layered structures and it does so simultaneously in all the layers. The method uses a modest optical photocurrent spectroscopy setup at ambient conditions. The results are analyzed using a simple model for electro-absorption. As an example, we apply the method to a typical GaN high electron mobility transistor structure. Measurements under various external electric fields allow us to experimentally construct band diagrams, not only at equilibrium, but also under any other working conditions of the device. The electric fields are then used to obtain the charge carrier density and mobility in the quantum well as a function of the gate voltage over the entire range of operating conditions of the device. The principles exemplified here may serve as guidelines for the development of methods for simultaneous characterization of all the layers in complex, multi-semiconductor structures.

## 1 Introduction

While silicon technology is facing the challenges of quantum mechanical tunneling, bandgap-engineered devices thrive on quantum mechanical effects to produce faster switching.[1] Yet, while silicon technology can do with a single semiconductor, bandgap engineered devices typically require a stack of several nanometer-scale thin-films of semiconductors of different bandgaps. These complex heterostructures present a challenge to silicon-era characterization tools, which are mostly capable of characterizing structures made of a single semiconductor material. Photoemission spectroscopies, e.g., x-ray and ultra-violet photoelectron spectroscopies, have been successfully used in studying band-structure of single heterojunctions,[2,3] but fall short of characterizing stacks of more than a single heterojunction, such as those in a typical light emitting diode, laser diode, or a high-electron-mobility transistor. So far, this shortage in characterization methods has been compensated for mostly by pure simulations, semi-empirical simulations, and theoretical calculations.[4,5,6]

The advantage of using junctions of more than a single semiconductor was already recognized by Shockley in his patent of the bipolar junction transistor, while the foundations for the use of semiconductor heterostructures were laid later by Kroemer,[7] and dubbed "bandgap engineering" by Capasso.[8,9] The early work on heterostructures at Bell Labs was also the ground for the invention of modulation doping by Dingle[10] followed by the invention of the high electron mobility transistor (HEMT).[11] The HEMT evolved from a single GaAs-AlGaAs heterojunction in 1980 to the state-of the-art of a multi-layer AlGaN-GaN of today.[12,13] One of the challenges faced by bandgap engineers today is to verify that the engineered band structure of the multi-layer heterostructure was actually accomplished. Layer thickness and composition are used to estimate the bandgap and band-offsets that should result, while a more complex process is required for the estimate of the built-in electric fields. Eventually, when the structure is materialized, many methods can be used to give partial validation of some of the parameters. In practice, however, the designer will rarely bother to use several complimentary methods to obtain a rough estimate of the band structure, and in most cases, will make do with the final electrical tests of the device.

The purpose of this work was to develop an experimental tool to measure the band structure: bandgaps, band offsets, and built-in electric fields, in a multi-layer heterostructure, at equilibrium, and also under external electric fields – all using a single method, and in a single measurement. Here, we propose a tool to characterize, simultaneously, all the layers in a multi-semiconductor structure and to construct energy band diagrams of complex heterostructures, not only at equilibrium, but also over the entire range of operating conditions. We demonstrate the method on a HEMT structure without limiting the generality, as we assume that the required test structure can always be fabricated on any semiconductor heterostructure.

## 2 Proposed Method and Model

The approach taken here is to use an optical spectroscopy in conjunction with an electrical measurement. While the most commonly used optical spectroscopy is photoluminescence, it will reveal, in most cases, only the lowest bandgap in the structure. This is because carriers tend to descend to the lowest bandgap before recombining. To avoid this limitation, one may use a spectroscopy that is based on *absorption* rather than *emission*. In a structure comprised of *nanometer-scale-thin* layers, photons can reach and be absorbed in any layer in the structure. The absorption of photons takes place mainly at the bandgap energy of each layer, and therefore, responses of layers of different bandgaps should be observed at different photon energies within the same spectral response curve. This way, all or most of the responses may be recorded in a single measurement of a spectral response curve.

The absorption of photons at the bandgap energy generates electron-hole pairs, and the availability of photo-generated carriers gives rise to changes in various electrical properties. In this work, we chose to detect *electric current*. This configuration is commonly used in *spectral photo-conductivity* and in *internal photoemission*.[14,15] In the latter method, the current flows perpendicular to a potential barrier formed between two materials, and only when the photon energy is great enough to excite electrons over the barrier, can an electric current be detected. Thus, in a stack of more than a single heterojunction, a current can be observed only at photon energies exceeding the highest barrier. This renders the internal photoemission configuration inadequate for our purposes.[16] A better configuration would be to fabricate two parallel contacts to the lowest bandgap in the structure. Optically generated carriers will typically descend to the lowest bandgap layer, and therefore, the responses of all the layers may be detected as steps in the electric current – a single step for each bandgap-energy in the structure. A spectrum taken from a typical HEMT device, showing a set of such steps, is shown in **Figure 1**. The layer structure of the device is shown in the inset. Such a spectrum contains all the possible band-to-band transitions, whether within the same material (the bandgap), or between adjacent materials. The energy differences between these two types of transitions can be used to calculate band offsets.

Each spectral step contains information not only on the optical transition energy (bandgap), but also on the electric field in the corresponding layer. Careful inspection reveals that each step commences well below the actual bandgap and rises in a sloped manner. This early photocurrent response reflects the shape of the absorption edge in semiconductors. The absorption edge, as reflected in the photocurrent, is affected mainly by the typically strong electric fields present at semiconductor interfaces.[17] The electric field assists photons with energy smaller than the bandgap to excite electrons across the forbidden gap by adding energy from the electric field. The effect of electric fields on the absorption edge in semiconductors is commonly known as the Franz-Keldysh effect.[18,19] We have previously modeled this effect on photoconductivity in single-material structures.[20] Using the same model on the data of each step produces a linear curve that intercepts the photon energy axis at the exact bandgap (or optical transition) energy, and the slope of this curve may be used to obtain the maximum electric field in that layer. Each spectrum is thus analyzed for its various steps to produce a band diagram.

The electrical current at each step of the photo-response was modelled using Eq. 1[20]

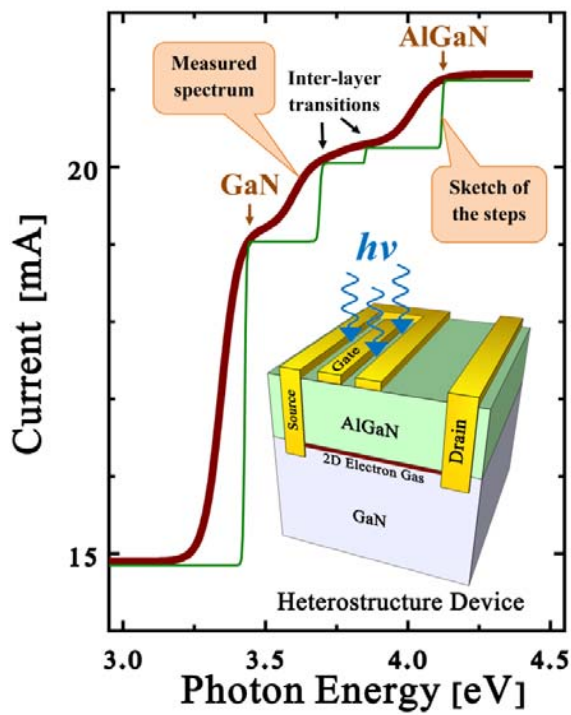

**Fig. 1** Typical photocurrent spectrum measured between the drain and source contacts as a function of photon energy (Red curve). The spectrum is comprised of a series of steps, each saturating around its corresponding optical transition energy. When an optical transition takes place between the valence band and the conduction band of the same material, the photon energy equals the bandgap of this material. The photocurrent rise associated with each transition precedes the energy of the actual transition due to an effect of the electric field in the corresponding layer. Thus, the preceding slope can be used to measure the electric field in the corresponding layer. We added a sketch of the steps without their preceding slope to aid the eye (green curve). The inset shows the structure of the high electron mobility transistor used in this experiment. The top 40 nm of such device typically contains several active layers of different semiconductors. Using the proposed method, it is possible to obtain for each layer the bandgap and the maximum electric field under any external electric field. It is also possible to get the band offsets from interlayer transitions.

$$I(h\nu) = I_D + (I_S - I_D)[1 - R(h\nu)]exp\left(-\left(\frac{E_g - h\nu}{\Delta E}\right)^{3/2}\right) \quad (1)$$

where ID – dark current (practically it is the current preceding the rise), IS – current following the rise, and R(hv) – spectral reflectance from the surface of the sample (in practice, we took into account only the reflection of the air interface, because we found the effect of reflection in general to be rather minor and negligible for inter- and intra-layer reflections that we neglected), Eg is the bandgap or the energy of the involved optical transition, hν is the photon energy, and ΔE is given by

$$\Delta E = \left(\frac{3}{4}\frac{qE\hbar}{\sqrt{2m}}\right)^{2/3} \quad (2)$$

where $E$ is the maximum of the electric field in the layer, $q$ is the electron charge, $m$ is the reduced effective mass, and $\hbar$ is the reduced Plank constant. Rearranging Equation 1, we get

$$y(h\nu) = \left[\ln\left(\frac{I_S - I_D}{I(h\nu) - I_D}\right) + \ln[1 - R(h\nu)]\right]^{2/3} = \frac{E_g - h\nu}{\Delta E} \quad (3)$$

The advantage of Eq. 3 is that its right-hand side is a linear expression of the photon energy. Presenting the data this way, each step response transforms into a linear curve that intercepts the photon energy axis at the exact optical transition energy. From the slope of this line, one can extract the *maximum electric field* in that semiconductor layer. The same treatment is given independently to each of the layers of a complex semiconductor heterostructure.

As a matter of fact, the basic method does not require more than *two Ohmic contacts* to construct the *equilibrium* band diagram. However, the use of a complete transistor affords an additional experimental handle, the gate. The gate can be used to apply an external electric field during the spectral acquisition. Using a set of spectra, acquired over a range of gate voltages, it is possible to construct a set of band diagrams for a range of operating conditions of the device.

**Figure 2** shows a surface plot of a set of 150 photo-current spectra obtained over a photon energy range of 2.9 to 4.425 eV (5-meV steps) and a gate voltage range of 0 to -7.5 Volts (in 50-mV increments) from the heterostructure described in Figure 1. The curves were obtained under source to drain voltage of 0.1 Volts – within the linear mode of the transistor. Similarly, we also applied the method for source-drain voltage of 8 Volt corresponding to the saturation mode of the transistor (not shown). The top panel of **Figure 3** shows overlapped plots of normalized photocurrent spectra at the photon-energy range near the bandgap of GaN. Each plot was acquired under a different external field. The plots were cut from the same spectra shown in Figure 2 to emphasize the effect of the applied external field on the various photocurrent steps – the slopes preceding the band-edge are observed to decrease with the increasing applied field. Applying the model (Equation 3) to each of the spectra on the top panel of Figure 3 produces a corresponding set of linear curves (Figure 3 – bottom panel). This graphic method provides an easy confirmation to validate the assumption of Franz-Keldysh electro-absorption. If the Franz-Keldysh effect does not take place, the use of Equation 3 is unlikely to produce a linear curve. All the curves intersect the photon energy axis at the GaN bandgap, while the slopes are observed to decrease with the increasing field. A similar analysis was also carried out for the step response of the AlGaN. The linear portion of each curve is observed over a range of about 0.1 eV preceding the bandgap. Below this range, the data deviate from the linear course, because our model approximates a parabolic barrier with a triangular barrier. This approximation is good close to the bandgap but departs from reality further away. At the bandgap, the spectra deviate again from the linear course due to the different above-gap physics.

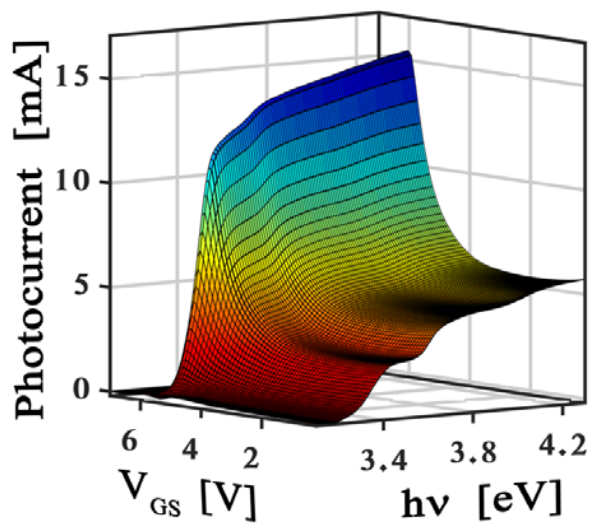

**Fig. 2** A series of 150 channel photocurrent spectra acquired from the high electron mobility transistor (HEMT) shown in Fig. 1 under a range of external electric fields. An external field was applied by applying voltage to the gate contact of the transistor ($V_{GS}$), while the voltage between the drain and the source ($V_{DS}$) was kept constant. In this figure, $V_{DS}$ was 0.1 V.

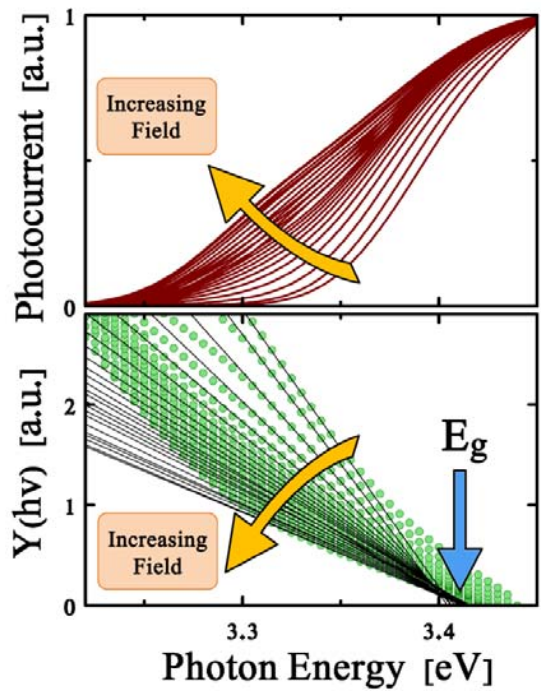

**Fig. 3** Top panel: Overlapped plots of photocurrent spectra at the photon-energy range near the bandgap of GaN. Each plot is for a different external field. The plots were cut from the spectra shown in Fig. 2 to show the effect of the applied external field on the GaN step response – the slope preceding the band-edge is seen to decrease with the increasing applied field. Bottom panel: A corresponding set of straight lines is obtained by applying the model (Eq. 3) to each of the spectra in the top panel. In both plots, the sampling has been reduced for clarity.

## 3 Experimental Details

The semiconductor heterostructure was grown by metal organic chemical vapor deposition (MOCVD) on c-plane sapphire. The layer sequence was an AlN nucleation layer, 2 µm of undoped GaN, 11 nm of Al30Ga70N, and 1.5 nm GaN cap layer. For device isolation, shallow mesas were dry etched in chlorine-based plasma. After removing the GaN cap layer, 100 nm of Si3N4 was deposited on top of the structure by plasma enhanced chemical vapor deposition (PECVD). Contact pads and 3 µm wide gate trenches were dry etched in Si3N4. All the metal contacts were deposited using e-beam thermal evaporation. Source and drain Ohmic contacts were Ti(30nm)/Al(70nm)/Ni(30nm)/Au(100nm) annealed at 900 °C for 1 min in nitrogen ambient. The gate contact was a Ni(30nm)/Au(100nm) Schottky barrier. For spectral data acquisition, the samples were placed in a dark and shielded box at atmospheric room temperature conditions. Illumination was applied from the gate side. For illumination, we used a 300 Watt Xe light source, monochromitized using a Newport Corp. double MS257 monochromator and further filtered by order-sorting long-pass filters. During spectral acquisition, a constant voltage was applied between the source and drain contacts. Electrical measurements were carried out using two Keithley 2400 source-meters. To avoid the effect of light on the measured electric field, we worked at a photon flux small enough, so that the maximum produced photocurrent is about two orders of magnitude smaller than the dark current. The intensity of light was 6.5 µW/cm2 at 280 nm. To avoid features resulting from the spectral distribution of the lamps, we operated the spectrometer in a closed control loop maintaining a constant photon flux throughout the spectral range of the measurement. The wavelength was stepped at equal photon energy steps. At each photon energy point, a full scan of the gate voltage range was performed. Each data point is an average of 30 consecutive measurements.

## 4 Results and Discussion

A full analysis of a typical transistor is given in **Fig. 4**. The figure has two columns. The left column shows an analysis at the linear mode, while the right column shows the same for the saturation mode. The first row shows the maximum electric fields in the AlGaN and GaN layers as a function of the applied gate voltage, as calculated from the optical response curves. The AlGaN layer is fully depleted, and therefore the field does not vary with the position within the layer. On the other hand, the GaN is in a state of accumulation at the heterojunction, forming a triangular quantum well with 2-dimensional electron gas (2DEG). The field in the GaN layer reaches its maximum within the quantum well and gradually decreases to zero as one gets away from the junction. The point where the optically induced band-to-band transitions take place in the layer is (always) the point where the electric field is the largest, provided there exist allowed states for electrons. For the GaN, this point is where the first discrete level (eigenstate) in the well meets the GaN conduction band.

Given the evolution of electric fields on both sides of the 2DEG, it is now possible to calculate its sheet charge density as a function of the gate voltage. For example, in our AlGaN/GaN structure, the 2DEG charge density, $qn_S$, is given by the discontinuity in the electric displacement field at the boundary between the AlGaN and GaN layers. The electric displacement field on the AlGaN side, $D_{AlGaN}$, is the sum of the known spontaneous polarization in AlGaN, $P_{SP,AlGaN}$, the piezoelectric polarization resulting from the mismatch to the GaN layer underneath it, $P_{PE,AlGaN}$, and the measured electric field, $E_{AlGaN}$. In the GaN, the electric displacement field outside the quantum well is created by $P_{SP,GaN}$ and the measured electric field, $E_{GaN}$.[21,22] The values of the polarization vectors are *-0.034 Cb/m2*, *-0.0464 Cb/m2* and *-0.00983 Cb/m2* for $P_{SP,GaN}$, $P_{SP,AlGaN}$ and $P_{PE,AlGaN}$ (Al composition of 30%), respectively.[23] Hence, we get[24,25]

$$qn_S = (\varepsilon \vec{E}_{AlGaN} + \vec{P}_{SP,AlGaN} + \vec{P}_{PE,AlGaN}) - (\varepsilon \vec{E}_{GaN} + \vec{P}_{SP,GaN}) \quad (4)$$

Substituting the measured electric fields in the AlGaN and GaN layers in Eq. 4, we can now graph the evolution of 2DEG charge density as a function of the gate voltage (second row in Fig. 4). Approaching the threshold voltage, the 2DEG gradually diminishes and the resistance of the channel increases. Since we measure the channel current, the photocurrent diminishes as well at this range, until, at a certain low value, the vanishing signal to noise ratio makes the results less reliable. This range is shaded in gray in the first row. We extrapolated the trend preceding this range and used this extrapolation in the calculation of the 2DEG charge density within the uncertain range.

The third row of Fig. 4 shows drain current measured in the dark over the same range of gate voltages in both modes. As expected, it roughly follows a similar trend as the 2DEG charge density. Using this drain current and the 2DEG charge density, we were now able to draw the channel mobility as a function of the gate voltage in each of the transistor modes. The mobility is shown at the bottom, fourth row, of Fig. 4.

Using the method on our AlGaN/GaN structure provided us with the bandgaps, band offsets, and built-in electric fields. This is basically all that is required to construct a band diagram. Since we get a band diagram for each of the applied gate voltages, we can actually draw the evolution of the band diagram with the applied external field. **Figure 5** shows 3 specific band diagrams for 3 specific gate voltages. To construct the band diagram, we assumed linear bands in the AlGaN layer. The GaN layer required a one-dimensional Poison-Schroedinger equation solver. The solution used our measured values of the 2DEG charge density and the GaN electric field at the first sub-band in the GaN quantum well.

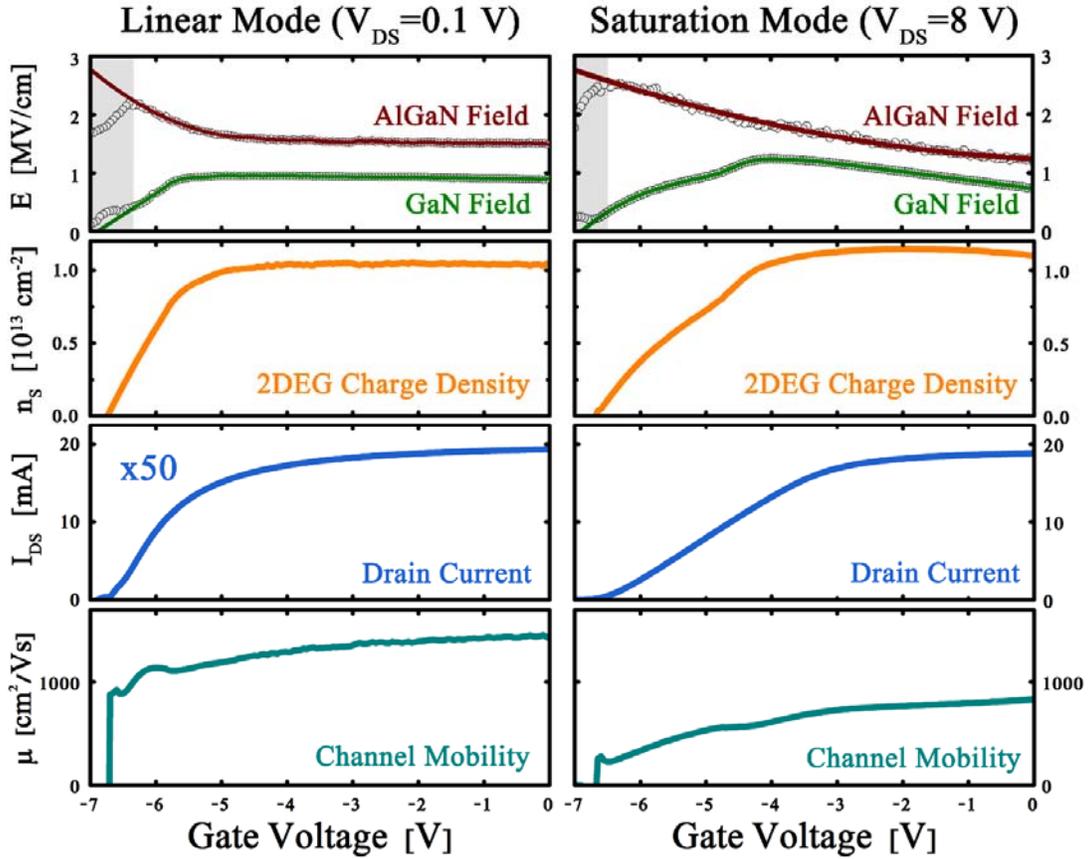

**Fig. 4** Two sets of data for two transistor modes shown as a function of the gate voltage: 1 Linear Mode – Left column figures, and 2 Saturation mode – Right column. <u>Row 1</u>: Peak (maximum) electric fields in the GaN and the AlGaN layers obtained from the measured spectra. Note that the measured field is the maximal value of the field also in the lateral dimension, i.e., along the channel. The maximal field along the channel occurs in this device under the gate, at its drain side. <u>Row 2</u>: 2DEG charge density calculated from the electric fields of Row 1. <u>Row 3</u>: Drain current measured in the dark. <u>Row 4</u>: Channel mobility calculated from the 2DEG charge density of Row 2 and the drain current of Row 3. Note the grayed regions in Row 1. At these regions the channel is almost closed, the drain current drops, and therefore the sensitivity of our method reaches a limit. The data in these ranges (open circles) is not reliable. We have extrapolated the data as shown in the full-line curves. The 2DEG charge densities over these ranges are calculated from the extrapolated curves. The same caveat goes to the channel mobility of Row 4. However, mobility is meaningless where no conduction is possible.

So far, we have treated only the case of electron-hole generation. Excitons are likely to be generated as well. Dow and Redfield showed that excitonic absorption followed a model different from the Franz-Keldysh model.[26] If excitonic absorption affected the photocurrent, our graphic method would not yield a straight line. Therefore, if we do obtain a straight line, it serves to confirm the adequacy of the Franz-Keldysh model. The absence of exciton expression in our photocurrent is probably due to the fact that excitons are electrically neutral and require significant dissociation to be able to contribute.

Dissociation of excitons is not always an effective process and poses a major bottleneck in photovoltaic efficiency.

Another case not addressed so far is the absorption in quantum wells and the adequacy of the Franz-Keldysh model to various cases falling under this category. Franz-Keldysh electro-absorption requires that at least one type of carrier will not be confined. Hence, in the case shown here, of a triangular quantum well, the model is clearly adequate, because only electrons are confined. However, there is still the case of a double heterostructure. It has been shown by Miller *et al*. that high electric fields give rise to the quantum confined Stark effect.[27] This effect relates to excitonic absorption affecting the exciton binding energy and increasing its survival. As excitons do not carry charge, the Stark effect may not have a direct effect on photocurrent. Yet, another work by Miller *et al.* suggests that even if the effect of excitons is altogether excluded, bulk-like Franz-Keldysh effect, i.e. the smearing of the absorption edge to low energies, cannot take place in narrow quantum-wells.[28,29] However, one should bear in mind that their calculations were carried out under the assumption of infinite energy barriers. Most of the practical quantum wells are actually very far from meeting this assumption. Furthermore, valence band wells are typically extremely shallow with the energy separation between eigenstates on the order of the phonon energy, kT. Clearly, these are not true eigenstates. No real hole confinement can take place in most of these real cases, and hence, it may not be unpractical to expect a bulk Franz-Keldysh photocurrent response in most of the real quantum wells at room temperature.

The following limitations should be kept in mind when using the proposed method. First, the method can yield only the magnitude of the electric field but not its direction. Second, there is a practical lower bound to the detectable field. This is because various other effects may cause broadening of the photoresponse resulting in a minor slope of the photocurrent step. The most obvious source for such broadening is, for example, the resolution of the spectrometer. In this work, the lowest detectable electric field was ~*0.1 MV/cm*. In any case, the fields in heterojunctions are typically much greater than this bound (see e.g., the electric fields in Fig. 4). In field-effect transistors, the electric field varies in the lateral direction (along the channel) as well. Since electro-absorption takes place at the point of strongest electric field, our band diagrams show a cross-section of the transistor under the gate at its drain side, which is the point of strongest field in the HEMT. Finally, in the case of a HEMT, very often, a metal field-plate is placed on top of the gate that may optically screen the gate and its vicinity. In this case, it may be difficult to illuminate the point of highest electric field.

## 5  Conclusion

The proposed method provides the energies of the various optical transitions, which can be used to evaluate the energy gaps at the various layers and the interlayer band offsets. It also provides the built-in electric fields at each of the layers, which can be used to construct the band diagram of the structure and the interface charges, e.g., the 2DEG charge density in quantum wells, and the channel mobility. It may also be used under external electric fields to explore the dynamic behavior of the band structure and of the interface charges and mobility. The model is based on the Franz-Keldysh effect, and it provides a graphical confirmation for the validity of this assumption. In this work, we have used the channel current as the measured electrical property. Various other electrical properties may also be measured using the same setup, and their physics may be used for the evaluation of various other semiconductor material properties in complex bandgap-engineered semiconductor layer-stacks.

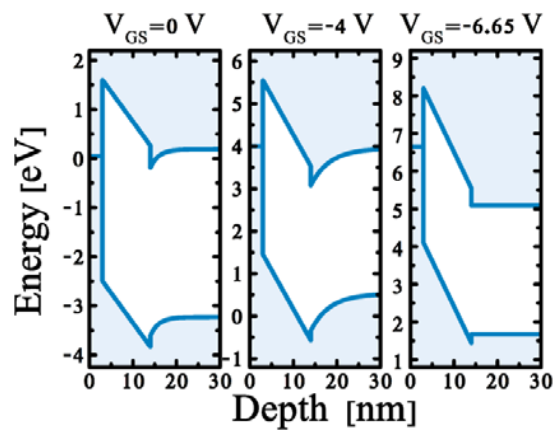

**Fig. 5** Band diagrams for 3 specific gate voltages (from zero to threshold voltage), calculated from the data of Fig. 4. The proposed method allows one to draw a band diagram for each single gate voltage (e.g., a diagram for each of the 150 spectra of Fig. 2) to visualize the evolution of the band diagram over the entire range of operation of the transistor.

## Acknowledgements

This work was funded by a research grant from the Israeli Ministry of Defence (MAFAT).